\documentclass[12pt,preprint]{aastex}
\usepackage{natbib}
\begin{document}

\title{Spatially Resolved Mid-IR Imaging of the SR 21 Transition Disk}

\author{J.A. Eisner\altaffilmark{1}, J.D. Monnier\altaffilmark{2}, 
P. Tuthill\altaffilmark{3}, S. Lacour\altaffilmark{4}}

\altaffiltext{1}{Steward Observatory, University of Arizona, Tucson, AZ 85721}
\altaffiltext{2}{University of Michigan Astronomy Department, 
941 Dennison Building, Ann Arbor, MI 48109}
\altaffiltext{3}{School of Physics, Sydney University, N.S.W. 2006, Australia}
\altaffiltext{4}{Observatoire de Paris, F-92195 Meudon, France}


\begin{abstract}
We present mid-IR observations from Gemini/TReCS
that spatially resolve the dust emission around
SR 21.  The protoplanetary disk around SR 21 is believed to have a cleared
gap extending from stellocentric radii of $\sim 0.5$ 
AU to $\sim 20$ AU, based on modeling
of the observed spectral energy distribution.  Our new observations resolve
the dust emission, and our data are inconsistent with the previous
model.  We require the disk to be completely cleared within $\sim 10$ AU,
without the hot inner disk spanning $\sim 0.25$--0.5 AU posited previously.
To fit the SED and mid-IR imaging data together, we propose a disk model
with a large inner hole, but with a warm companion--possibly surrounded
by circumstellar material of its own--residing near the
outer edge of the cleared region.  We also discuss a model with a narrow
ring included in a large cleared inner disk region, and argue that it is
difficult to reconcile with the data.
\end{abstract}

\section{Introduction \label{sec:intro}}
While young ($\la 1$ Myr) stars are generally surrounded by dust and gas 
rich protoplanetary disks \citep[e.g.,][]{KS95,MO96,BECKWITH+90}, 
these disks tend to disappear within a few million years 
\citep[e.g.,][]{HLL01a,EISNER+08}.  The mechanism by which protoplanetary disks
are dissipated remains unclear, although several processes--including
planet formation, viscous accretion onto the central star, or
photo-evaporation from stellar UV and X-ray photons--may play a role
\citep[e.g.,][]{CALVET+02,ACP06,NSM07}.  In any case, one expects
that as protoplanetary disks evolve, they will at some point transition
from being optically thick accretion disks to tenuous, optically thin disks.

Transition disks were first defined empirically based on observed
spectral energy distributions \citep[e.g.,][]{STROM+89}.  These objects
display long-wavelength ($\ga 10$ $\mu$m) infrared excesses indicative
of optically thick disks beyond stellocentric radii of a few AU.
In contrast to typical  young, optically thick protoplanetary disks,
however, transition disks lack opacity in the near-IR and produce
less near-IR excess emission.  This opacity hole is often interpreted as a
region in the inner disk that is largely cleared of material.

Several systems have now been modeled in terms of this cleared inner disk
picture \citep[e.g.,][]{CALVET+02,DALESSIO+05}.  In addition, recent studies
have found evidence that some objects appear to have hot inner disk material
and optically thick outer disks, but a cleared gap at intermediate radii
\citep[e.g.,][]{BROWN+07,ESPAILLAT+08}.  Most of these studies are based
on modeling of spectral energy distributions, although there are now
a few examples where the outer edges of cleared disk regions have been
imaged \citep[e.g.,][]{GOTO+06,RATZKA+07,HUGHES+07,BROWN+08}.  In addition,
there are now observations of optically thin material within the ``cleared''
regions in some systems \citep[e.g.,][]{ECH06}.

SR 21 is a young star in the Ophiuchus star forming region, at an 
assumed distance of 160 pc.  A spectral
energy distribution measured from $\sim 1$--$100$ $\mu$m
shows strong mid-IR emission indicative of a disk of material that
re-processes stellar emission, but a deficit of emission around 5 $\mu$m.
These data have been interpreted successfully with a disk model that extends
from stellocentric radii of 0.25 AU to 300 AU, but with a cleared gap
between 0.45 and 18 AU \citep{BROWN+07}.  Observations of emission from
ro-vibrational transitions of CO show evidence for a narrow ring of
gas near a stellocentric radius of $\sim 7$ AU \citep{PONTOPPIDAN+08},
implying a substantial column of gas within the putative cleared region
evinced by the SED modeling. 

Detection of broad, but weak H$\alpha$ emission suggests accretion of some
disk material onto the central star \citep{PAJ08}, although non-detection
of Pa$\beta$ emission suggests that the accretion rate is $\la 10^{-9}$
M$_{\odot}$ yr$^{-1}$ \citep{NTR06}.  The dust mass of the SR 21 disk 
has been inferred to be $\sim 2 \times 10^{-4}$ M$_{\odot}$
\citep{AW07,PAJ08}, on the high end of the disk mass distribution for T Tauri 
stars \citep[e.g.,][]{EISNER+08}.
Thus, while the source appears to have
a fairly massive outer disk, the accretion rate is lower than that of a
typical classical T Tauri star, as found for other sources classified
as transition objects \citep{NSM07}.

Here we present spatially resolved mid-IR observations of SR 21.  Mid-IR
observations are sensitive to scales near the outer edge of the cleared region,
or even within the cleared region, and so can critically test the 
geometric model inferred from the spectral energy distribution.  Furthermore,
such observations can potentially reveal structures that might be associated
with planet formation or other disk clearing processes.

\section{Observations and Data Reduction \label{sec:ob}}
We observed SR 21 and several bright, unresolved calibrator stars
at 8.8 $\mu$m (Si-2 filter) and 11.6 $\mu$m (Si-5 filter) 
wavelengths with the TReCS
camera on the Gemini South telescope on UT 2007 May 9.  To obtain diffraction
limited images, we employed very short integration times
to effectively freeze the atmosphere.  For each source at each wavelength, we 
obtained 39 172-ms integrations for two sets of two up/down dither positions
and two left/right nods; the total is 312 integrations per source per 
wavelength.

For each observed target, the data from adjacent nod positions were subtracted 
in order to remove the sky background.  This procedure appears to work 
reasonably well, leaving $<300$ counts RMS residual, 
compared to peak target fluxes of $\sim 50,000$ counts.

We analyzed the calibrated images using techniques borrowed from non-redundant
aperture masking interferometry.  
A ``pseudo-mask'' consisting of 27 sub-apertures was created
and we computed the visibilities for each of the 351 baselines.  Because
our aperture is not, in fact, non-redundant (i.e., each of our visibilities
contains contributions from a number of identical baselines), each measured
visibility includes redundancy noise.  To calibrate this effect
we computed the visibilities for a number of unresolved ``check stars.''
Uncertainties are largest for intermediate baselines, where
photon counts are lower than for the shortest baselines but
redundancy noise is higher than for the longest baselines.

We azimuthally average the 2-D visibility distributions, to enhance effective
signal-to-noise and for ease of plotting.  We typically use these
azimuthally-averaged visibilities in the analysis below.


%


\section{Modeling \label{sec:mods}}
We begin this section with
a simple estimate of the size scale of the mid-IR emission seen in our
observations.  We then reproduce the gapped-disk model of \citet{BROWN+07}
and demonstrate that, while it fits the
observed SED well, it does not provide a good fit to our imaging data.
We then generate a new grid of models and attempt to obtain a better 
fit to the combined SED+mid-IR imaging dataset.

\subsection{Estimated Size of the Mid-IR Emission \label{sec:mirsize}}
To provide a rough estimate of the size scale over which the mid-IR
emission seen in our observations is distributed, we fit simple,
geometric ring models to the visibilities \citep[e.g.,][]{EISNER+04}.  
We fit the data at 8.8 $\mu$m
and at 11.6 $\mu$m separately.  At 8.8 $\mu$m, the visibilities are
fitted well with a ring model with a radius of $11 \pm 1$ AU
($67 \pm 5$ mas).  
At 11.6 $\mu$m, the best-fit ring model has a radius of $15 \pm 1$ AU
($92 \pm 5$ mas).
The data at each wavelength are fitted well with
these simple models, and the fits provide reduced $\chi^2$ values 
well below unity.  

The emission at 8.8 $\mu$m appears
significantly more compact that the emission at 11.6 $\mu$m.
This is due, in part, to the larger stellar contribution to the
emission at shorter wavelengths.  However, if we use stellar properties from
the literature \citep[e.g.,][]{BROWN+07} and account for the
stellar flux at each wavelength, we still find a significantly
larger angular extent of the emission at 11.6 $\mu$m compared to
8.8 $\mu$m.  The larger size at longer wavelength implies a temperature 
gradient with cooler material distributed more widely than hot 
material.  

We also fitted an inclined ring model \citep[e.g.,][]{EISNER+04} to 
the 2-D visibilities in order to explore potential asymmetry in the data.
At 8.8 $\mu$m, the best-fit model has an inclination of 
$26^{\circ} \pm 6^{\circ}$ and a position angle (PA, east of north) of
$44^{\circ} \pm 7^{\circ}$.  At 11.6 $\mu$m, the inclination is
$27^{\circ} \pm 3^{\circ}$ and the PA is $61^{\circ} \pm 8^{\circ}$.


\subsection{Testing the gapped-disk model \label{sec:brownmod}}
\citet{BROWN+07} compiled an SED for SR 21 from $\sim 1$--100 $\mu$m,
and modeled it with a disk including hot material from 0.25 to 0.45 AU
that reproduced the observed near-IR excess,  a cleared gap extending from 
0.45 to 18 AU, and cool outer disk material at larger stellocentric radii
that reproduces the large far-IR luminosity of the object.  

Here we reproduce the model of \citet{BROWN+07}.  We maintain all of their
assumed model parameters, except for the inclusion of 
mm-sized grains in the mid-plane.  Since we are concerned
only with the region of the disk in the vicinity of the gap 
we may safely ignore these large grains.  We also assumed a blackbody for
the stellar emission, rather than the Kurucz model used by \citet{BROWN+07}.
While \citet{BROWN+07} do not specify the
radial dependence of the disk surface density assumed in their modeling,
we set this as $\Sigma(R) \propto R^{-1}$.
We computed an SED and images at 8.8 
and 11.6 $\mu$m for this model using the Monte Carlo radiative transfer code
RADMC \citep{DD04} without vertical structure iteration.

\begin{figure}[tbh]
\plotone{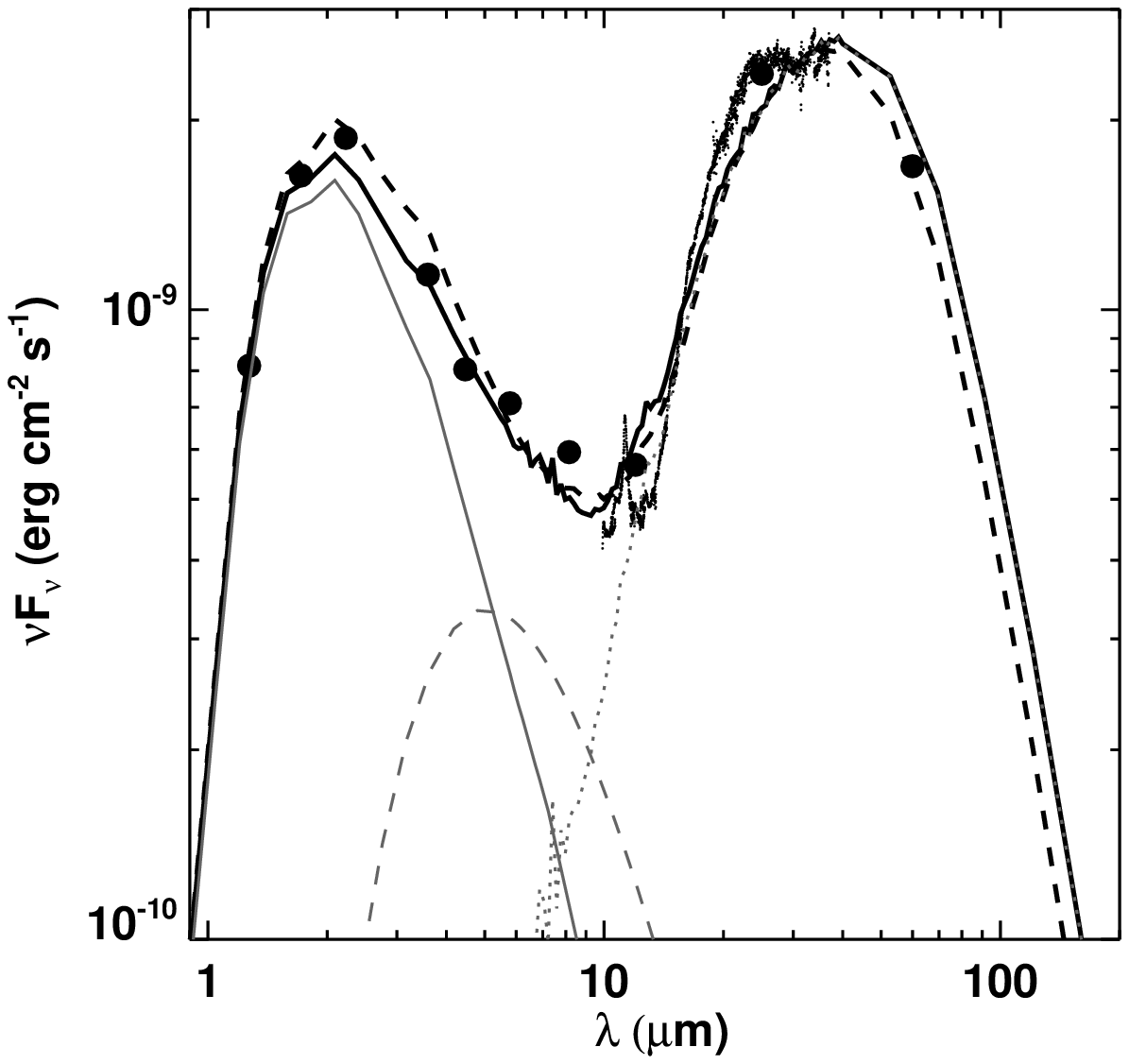}
\caption{Observed SED for SR 21 (broadband photometry: {\it large points};
IRS spectrum: {\it small points}) with model predictions 
overplotted.  We show the SED for a model that nearly matches the one 
derived previously by \citet{BROWN+07} ({\it dashed black curve}), as well
as for the disk+companion model described in \S \ref{sec:companion} 
({\it solid black curve}).  The stellar contribution to both models
is shown with the solid gray curve.  The contribution from the companion
and from the outer disk in the disk+companion model are shown with a dashed 
gray curve and with a dotted gray curve, respectively.  The
parameters of these models are given in the text.
\label{fig:sedfit}}
\end{figure}

\begin{figure}[tbh]
\plotone{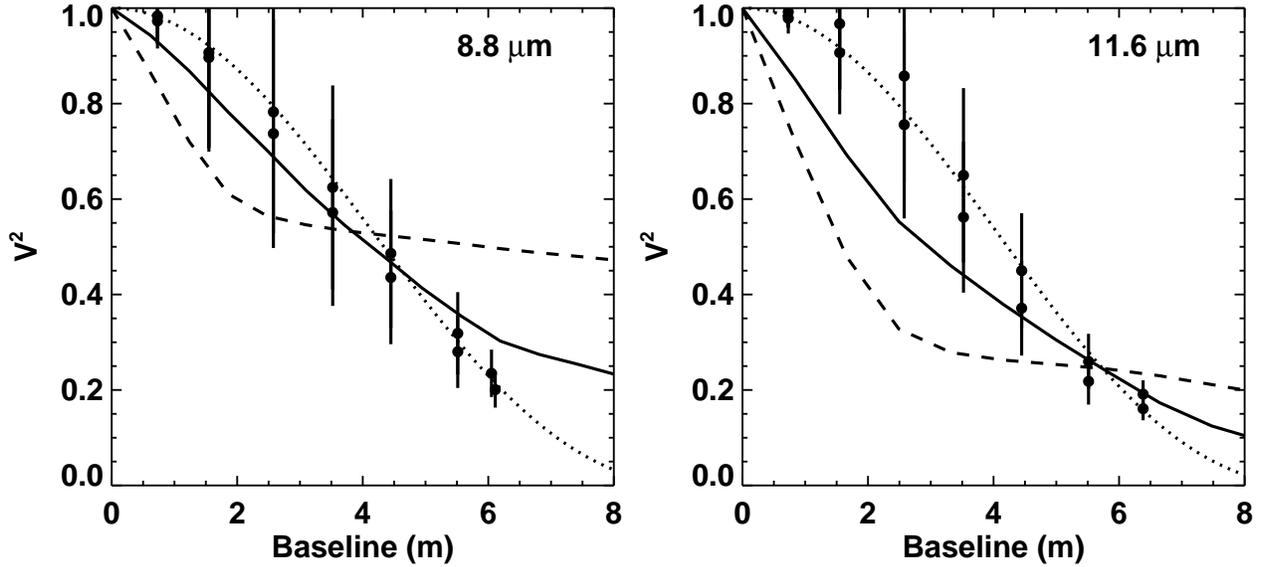}
\caption{Mid-IR squared visibilities computed from our 
short-exposure images
from TReCS ({\it points}) with the predictions of several models.
The two sets of points represent two datasets taken $\sim 2$ minutes apart.
The visibilities predicted by the SED-based, gapped
disk model of \citet{BROWN+07} are shown with the dashed curve.  The solid
curve shows the visibilities expected for the disk+companion model
described in \S \ref{sec:companion}; this model was
fitted to the mid-IR visibility data and SED data simultaneously.  We also
show the visibilities for simple ring models that were fitted separately to the
visibility data at each observed wavelength (\S \ref{sec:mirsize}).
The observed
and predicted visibilities plotted here have all been azimuthally averaged.
\label{fig:visfit}}
\end{figure}

The resultant model SED is shown in Figure \ref{fig:sedfit}, along with the
SED data from \citet{BROWN+07}.  We are able to reproduce the
previous model, and fit the SED well.  However, this model does not fit our 
mid-IR imaging data.  Figure 
\ref{fig:visfit} shows the visibilities predicted by this model at
8.8 and 11.6 $\mu$m.  These visibilities were computed from model images
using the same procedure applied to the
data.   

The model visibilities are clearly different than the 
observed visibilities.  In particular, the model has a strong compact
component that produces higher visibilities (with respect to the data)
at the longest baselines, and an extended component that produces lower
visibilities at shorter baselines.  The compact component arises from
the hot inner disk in the model, which extends from $\sim 0.25$ to 0.45 AU.
The extended component corresponds to 
the warm inner edge of the (outer) disk and the flared outer regions.
The fact that the shape of the observed visibility curves resemble 
those of single-component models with a size scale of $\sim 10$--20 AU
suggests that the hot inner disk and some of the flared outer disk
may need to be eliminated
from the model before it can be made consistent with the data.

\subsection{Disk Models \label{sec:diskmods}}
We attempted to fit a variety of disk models to our data, generating $\sim 500$
models using RADMC.  We varied the inner and outer disk 
radii, the inner and outer radii of the cleared gap, the factor by which
the density in the gap is reduced, the disk mass, the power-law indices 
describing the disk surface density and flaring profiles, and the disk scale 
height.  This model grid included disks that had neither gaps nor 
large clearings, disks with gaps, and disks
with large inner clearings.  We attempted to minimize the $\chi^2$ residual
between these models and our combined mid-IR imaging+SED dataset.
We assumed the SED data were uncertain by 
5\%, and used the uncertainties in the visibilities estimated from our 
reduction procedure (and plotted in Figure \ref{fig:visfit}). 

None of the models considered provided a substantially better fit than
the gapped-disk model discussed in \S \ref{sec:brownmod}.  This is not
surprising in light of the arguments given above.  Disk models that 
include hot emission at small stellocentric radii ($\sim 0.25$ AU) 
produce mid-IR emission much more compact than observed.  But disk models
lacking this hot, compact material, produce substantially less flux
from $\sim 2$--5 $\mu$m than observed in the SED.

\subsection{Disk + Companion Model \label{sec:companion}}
Here we attempt to reconcile the observation that the SR 21 system does appear
to have flux in excess of the stellar photosphere at near- to mid-IR
wavelengths (as seen in the SED) with the inference (based on the mid-IR
imaging data) that this flux cannot arise from the sub-AU scales expected 
for the inner disk.  Our proposed solution is that
this near- and mid-IR emission traces a warm ``companion'' located 
near the outer edge of the cleared region. Given the youth of the SR 21 system,
the companion may include emission from both (sub-)stellar and 
circum(sub)stellar emission.  We show below that
as long as the emission from the companion is marginally 
resolved in our mid-IR 
imaging data (i.e., if it has a stellocentric radius of $\sim 10$--20 AU),
the companion emission does not significantly affect the computed visibility
curve, and 
a disk+companion model can approximately reproduce the available data.

We implemented this model by first generating a grid of models with large
inner clearings or gaps using the RADMC code (see \S \ref{sec:diskmods}).  
We then added a companion to the 
model disk SEDs and images.  We generated a small grid of values for the 
temperature, radius, and stellocentric radius of the companion, and generated
models for each combination of these parameter values.  We placed 
the companion at a position angle of 50$^{\circ}$, following the direction
of asymmetry seen in the data (\S \ref{sec:mirsize}), although here we
azimuthally average the model visibilities and so
the chosen PA is not important.  
The main constraints on the temperature and radius
of the companion come from the near- and mid-IR excess fluxes, while
the stellocentric radius of the companion is constrained (weakly) by the
shape of the measured visibility curves. 

This model is not fully self-consistent, since the heating from 
the companion is not included in the calculation of the disk temperature 
structure.  Because the bolometric luminosity of the companion is small 
relative to the primary star, the additional heating that we neglect should
also be relatively small.  We are unable to determine rigorous error
intervals for our fits because we can only consider a limited number
of values for each parameter in order to keep the considered 
multi-dimensional parameter
space within the realm of computational feasibility.  Given
the uncertainties in the data, we estimate parameters are 
uncertain by at least 10\%.

The predicted SED and visibilities for 
our best-fitting model are shown in Figures 
\ref{fig:sedfit}--\ref{fig:visfit}.
and we show a synthetic image 
at 11.6 $\mu$m in Figure \ref{fig:modim11}.
The disk is cleared all the way into the central star, and this cleared
region extends outward to 15 AU.  
The disk mass is $10^{-5}$ M$_{\odot}$.
This disk has a scale height of 0.10 AU, smaller
than in \citet{BROWN+07}.  All of the other disk parameters are the
same as for the model in \S \ref{sec:brownmod}.
The best-fit model has a companion with a temperature of 730 K, a radius of 
40 R$_{\odot}$, and a stellocentric radius of 18 AU.
While the best-fit model places the
companion slightly outside of the cleared region, within uncertainties we
can not rule out the possibility that the companion resides within the
clearing.

The best-fit model produces asymmetric emission because of the companion.  The 
slight asymmetry seen in the mid-IR imaging data (\S \ref{sec:mirsize})
may reflect this companion flux.  The 
position angle derived from our data is different (at the $\sim 3\sigma$ 
level) from
the disk position angle inferred from observations of CO line emission
\citep{PONTOPPIDAN+08}, suggesting that the asymmetry in our mid-IR data
may trace a non-disk (e.g., companion) component.  However, the asymmetry seen
in the data is smaller than we would expect from our model.  If we fit
the 2D visibility data directly, the preferred value for the companion's
stellocentric radius is $\sim 7$ AU.  Given the noise in the data,
especially for the 2D visibilities where we do not have the benefit
of azimuthal averaging, we can not strongly constrain the position of the 
putative companion.

\subsection{Disk + Narrow Ring}
Another potential explanation of the data is a model including a large inner 
clearing with a narrow ring of emission at a stellocentric radius of 
$\sim 1$ AU (where disk temperatures are $\sim 700$ K).  This ring could add 
mid-IR flux without substantially altering 
the visibilities, similar to the companion discussed above.
We found, however, that such a model did not reproduce the data well.
Putting the emission at about 1 AU led to more compact emission than the 
companion model above, and hence a worse fit to the visibilities. Furthermore,
the inclusion of sufficient material to produce the correct level of flux 
excess led to a shadowed, and hence cooler, outer disk.  The combination
of these two effects led to models that do not provide good fits to the 
combined SED and mid-IR imaging dataset.  If one forced the inner ring to lie 
in a geometrically thin configuration, or allowed a substantial warp between 
the inner ring and the outer disk, the models could potentially be
made to fit better. It is unclear, however, whether such
modifications are physically realistic.

\section{Summary and Discussion}
The model used to fit the SED of SR 21, computed by \citet{BROWN+07}, is
not consistent with the mid-IR imaging data we obtained with TReCS.  In
\S \ref{sec:companion}, we argued that the SED and the mid-IR imaging
data could all be explained in the context of a disk model with a large
inner clearing (as opposed to a gap), and with a warm companion residing
near the outer edge of the cleared region.

Planetary or proto-planetary objects residing in the inner regions
of transition disks have been proposed as a potential cause for their
inner clearings \citep[e.g.,][]{GT82,BRYDEN+99,RICE+03}.  
Other sources classified as transitional disks have turned out to be
circumbinary disks \citep{GUENTHER+07,IK08}, 
where the outer disk may be truncated by tidal interactions
with stellar-mass companions \citep[e.g.,][]{LP79}.  

\begin{figure}[tbh]
\plotone{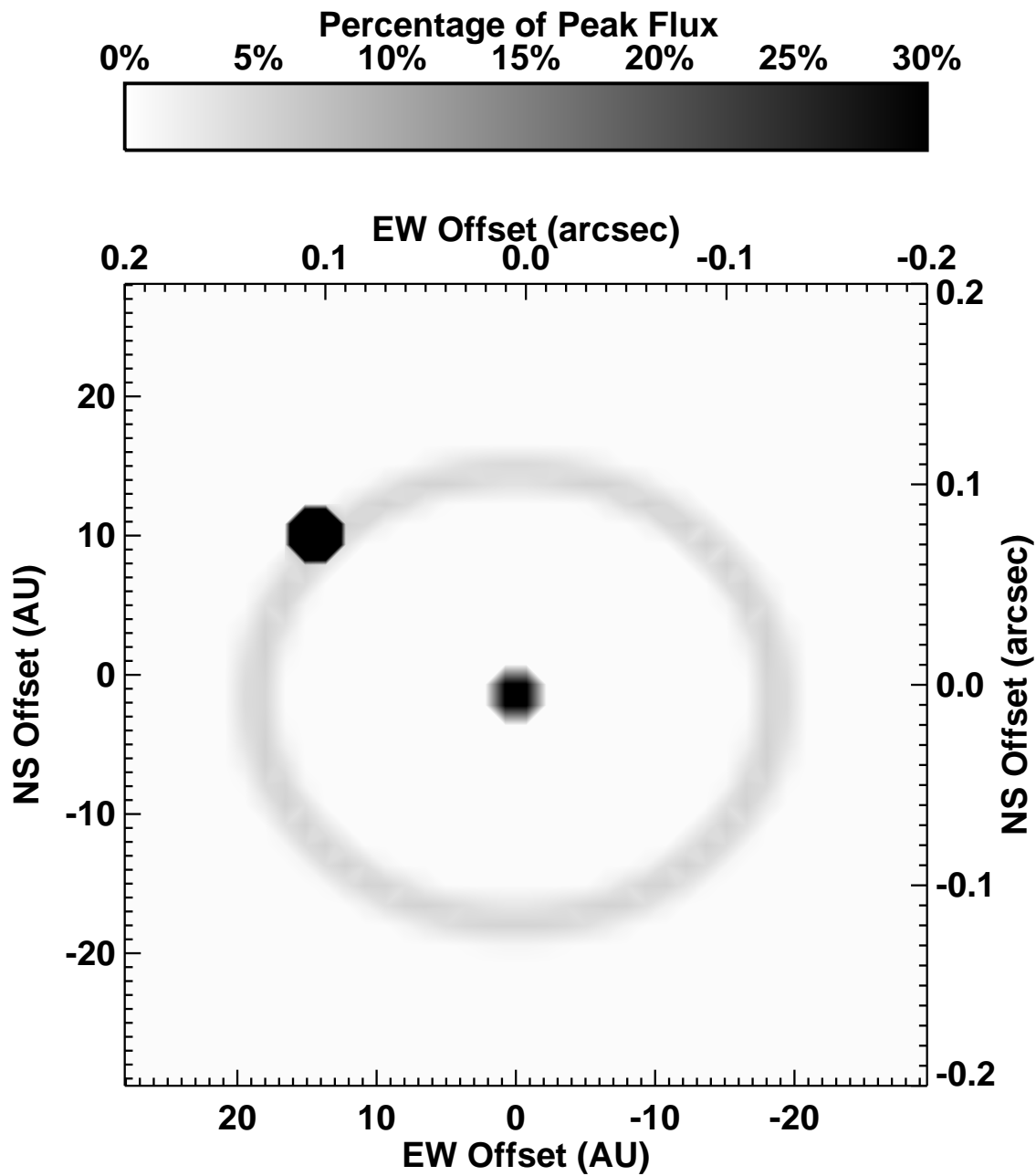}
\caption{A synthetic image of our disk+companion model (\S \ref{sec:companion})
at 11.6 $\mu$m wavelength.
We show only the inner 100 AU x 100 AU of the images, to highlight the 
structure near the cleared inner region.  Note that the exact location
of the companion is not well-constrained, and while the best-fit model
puts the companion within the outer disk, a location well within the cleared
region is compatible with our data.
\label{fig:modim11}}
\end{figure}

Given the inferred temperature and radius of the companion in our model, it is
unlikely to be a fully-formed star or planet.
However, an accreting object could be 
quite warm and large as it radiates away the copious energy liberated by the
infalling material. The ``companion'' in our model may correspond to the
cloud of accreting material around a forming proto-planet or 
low-mass protostar.
We can not constrain the stellocentric radius of the companion in our model
well enough to determine
whether it lies in the cleared region or in the innermost regions of the
disk.  If it actually resides in the disk, 
(as pictured in Figure \ref{fig:modim11}), 
its large inferred temperature
still lends credence to the idea that it is a self-luminous object, 
perhaps fueled by accreting material. 


We argued above that a disk model with a large inner clearing that contains
ring-like emission within the cleared region is difficult to reconcile with
the data. However, if a mechanism could be found that would prevent the inner 
disk material from casting a shadow over the outer disk--for example, if 
the matter was confined to a vertically-thin or highly inclined ring--then 
such a model could 
potentially fit the data.  The inference of a ring 
of gas near a stellocentric radius of $\sim 7$ AU  \citep{PONTOPPIDAN+08}
provides some support for such a model, although this gaseous ring
could also be explained in the context of sculpting by a companion.

High angular resolution imaging of SR 21 at shorter wavelengths is needed to 
further elucidate the nature of the emission within the ``cleared'' 
inner disk region.  Such observations can have higher resolution than those
presented here and would be more sensitive to the warm companion (or other
matter) inferred to reside in the inner disk.  Our models 
predict that in the $K$-band the companion 
has $\sim 5$\% the flux
of the central star.  In the $L$-band, the companion has $\sim 25\%$ the
flux.  At still longer wavelengths, the brightness ratio between the
companion and the star increases, but the disk also
brightens considerably, making detection of the companion more difficult.
The inferred stellocentric radius of the companion
places it $\la 100$ mas away from the central star, although our constraints
on the stellocentric radius are weak and it could be closer to the star.
The companion posited in our model may be detectable in near-IR ($K$ and $L$) 
adaptive optics observations, providing a potential test of the models
presented in this paper.



\medskip
\noindent
Based on observations obtained at the Gemini Observatory.
We are grateful to Kevin Volk, Tom Hayward, Adwin Boogert, Chris Packham,
and Charles Telesco for making these observations possible.  We also
thank Kees Dullemond for providing us with the RADMC code, and helping us
to get it running.  JDM acknowledges support from NASA-Origins NNG05G180G and
NSF-AST0352723.

\clearpage

\epsscale{1.0}

\epsscale{1.0}

\epsscale{0.9}





\end{document}